\documentclass[preprint]{aastex}

\usepackage{longtable}
\usepackage{graphicx}
\usepackage{amsmath,amssymb}
\usepackage{color}
\usepackage{units}
\usepackage{epstopdf}
\usepackage{hyperref}
\usepackage{multirow}

\newcommand{\beq}{\begin{equation}}
\newcommand{\eeq}{\end{equation}}
\newcommand{\bdm}{\begin{displaymath}}
\newcommand{\edm}{\end{displaymath}}

\newcommand{\SNR}{\textnormal{SNR}}

\definecolor{Gray}{gray}{0.9}

\graphicspath{{./plots/}}
\begin{document}


\title{Maximizing the Probability of Detecting an Electromagnetic Counterpart of Gravitational-wave Events}
\author{Michael Coughlin}
\affil{Department of Physics, Harvard University, Cambridge, MA 02138, USA}

\author{Christopher Stubbs}
\affil{Department of Physics, Harvard University, Cambridge, MA 02138, USA\\
Department of Astronomy, Harvard University, Cambridge MA 02138, USA}

\begin{abstract}
Compact binary coalescences are a promising source of gravitational waves for second-generation interferometric gravitational-wave detectors such as advanced LIGO and advanced Virgo.
These are among the most promising sources for joint detection of electromagnetic (EM) and gravitational-wave (GW) emission.
To maximize the science performed with these objects, it is essential to undertake a followup observing strategy that maximizes the likelihood of detecting the EM counterpart.
We present a follow-up strategy that maximizes the counterpart detection probability, given a fixed investment of telescope time.
We show how the prior assumption on the luminosity function of the electro-magnetic counterpart impacts the optimized followup strategy. Our results suggest that if the goal is to detect an EM counterpart from among a succession of GW triggers, the optimal strategy is to perform long integrations in the highest likelihood regions, with a time investment that is proportional to the $2/3$ power of the surface density of the GW location probability on the sky. 
In the future, this analysis framework will benefit significantly from the 3-dimensional localization probability.
\end{abstract}

\maketitle

\section{Introduction}\label{sec:Intro}
With the recent discovery of a compact binary black hole system \citep{AbEA2016a}, there is significant interest in the combined observation of electromagnetic (EM) and gravitational-wave (GW) emission \citep{AbEA2016b}. 
EM emission likely occurs on a variety of timescales and wavelengths ranging from seconds to months
in X-ray to radio, respectively \citep{Nakar2007,MeBe2012}. It is suspected that compact binary coalescences (CBCs) are also the progenitors of some or all short,
hard $\gamma$-ray bursts \citep{TrKi2008}. Plausible CBC event rates suggest that Advanced LIGO \citep{aLIGO} and Advanced Virgo \citep{adVirgo} could detect about 40 binary neutron star and 10 neutron star-black hole events per year of observation time \citep{AbEA2010}. In addition to CBCs, there are other possible sources of coincident EM and GW emission, including asymmetrical type II supernovae, soft $\gamma$ repeaters, anomalous X-ray pulsars, neutron stars recovering from pulsar glitches, cosmic string
cusps, or radio bursts.
Kilonoave are one promising source that can be identified, produced during the merger of binary neutron stars or a neutron star-black hole systems, likely peaking in the near-infrared with luminosities $\approx 10^{40}-10^{41}$\,ergs/s and lasting over a week \citep{KilonovaPrecursor,BaKa2013}.  

On the gravitational-wave side, a number of algorithms exist to derive inferences of compact binary source parameters based on gravitational-wave observations, as determined by Bayestar or LALInference \citep{SiPr2014,BeMa2015}. 
Bayestar is an algorithm which takes in information from compact binary search pipelines and returns gravitational-wave skymaps within seconds. 
LALInference instead provides inferences of intrinsic source parameters such as masses and spins, as well as extrinsic parameters such as sky direction and distance, but takes orders of magnitude longer to run.
These algorithms produce GW likelihood sky areas typically spanning $\approx 100\,\textrm{deg}^2$  \citep{Fair2009,Fair2011,Grover:2013,WeCh2010,SiAy2014,SiPr2014,BeMa2015}.
Finally, there are very-low latency algorithms proposed for performing rapid sky localization \citep{ChHo2015}.
There also exist algorithms to characterize generic gravitational-wave transients \citep{EsVi2015,CoLi2015}.

There has been a significant amount of work in recent years to improve follow-up of gravitational-wave sources with optical telescopes. Galaxy catalogs, such as the Gravitational Wave Galaxy Catalogue (GWGC) \citep{WhDa2011}, the Compact Binary Coalescence Galaxy Catalog \citep{KoHa2008}, and the 2MASS Photometric Redshift catalog \citep{BiJa2014} have been used to identify individual galaxies within the anticipated GW detection range. Another option is to rapidly create galaxy catalogs on-the-fly, after a gravitational-wave detection has been made \citep{BaCr2015}. Techniques for optimizing multiple telescope pointings also exist \citep{SiPr2012}. Antolini and Keyl \citep{AnHe2016} showed how to use the 2MASS Photometric Redshift catalog to optimize telescope pointings.

There are many factors that go into the probability of detecting a transient with a telescope. These include internal factors such as the exposure time, filter, field of view, and limiting magnitude, and external factors such as seeing and sky conditions. Fields of view are often approximately rectangular, although dead or defective pixels or vignetting from the optics can make this more complicated. 
We seek to maximize the likelihood of detecting an electromagnetic counterpart for a fixed allocation of observing time. This is especially important in an era with very different exposure times, limiting magnitudes, and field of views for telescopes. 
For example, the Dark Energy Camera on the Blanco 4m telescope at CTIO has a $3\textrm{deg}^2$ FOV, and 3600\,s r-band exposure length to reach 26\,mag in 1 arcsecond seeing \citep{DECamETC}, while LSST will have a $9.6\textrm{deg}^2$ FOV, and 810\,s r-band exposure length to reach the same \citep{LSSTETC}.
There are few operating or planned deep survey telescopes that have fields of view (FOVs) comparable to error regions of the gravitational-wave skymaps.
For example, the James Webb Space Telescope, a highly sensitive infrared space telescope with an expected launch in 2018, will have a $0.0013\textrm{deg}^2$ FOV \citep{BaHu2015}.
On the other hand, there are several shallow and wide-field operating telescopes such as Pi of the Sky \citep{MaBa2015}.

In this paper, we explore the benefits of optimizing single telescope pointings given limited time on the telescope. 
We show how adopting priors on the rate of compact binary detections and the distribution of sky areas produced by gravitational-wave detectors allows for a significantly more efficient follow-up than a naive follow-up strategy.
We will explore four particular cases, corresponding to two different mass distribution assumptions and two different prior flux assumptions.
The first assumes that in a particular field, there are 0, 1 or a few galaxies.
In this regime, the mass distribution will be very field-dependent.
In the other regime, there are many galaxies such that the mass distribution is no longer field dependent.
We will also explore two different luminosity distribution assumptions.
The first is a delta function prior, while the second is a flat prior.

We derive the scaling relations for optimizing telescope followup, in particular that the optimal exposure time allocated to any given field, under certain assumptions, can go as $t_i \propto \left(\frac{L_\textrm{GW}(\alpha_i,\delta_i)}{a(\alpha_i,\delta_i)}\right)^{2/3}$, where $L_\textrm{GW}(\alpha_i,\delta_i)$ is the gravitational-wave likelihood and $a(\alpha_i,\delta_i)$ is Galactic extinction.
This fits into a framework for planning optimal follow-up of gravitational-wave candidates.
We show that the required time to achieve a 90\% confidence level of detecting a gravitational-wave electromagnetic event is decreased by a factor of 3.
In this work, we will ignore a number of complications. One is the ``needle in the haystick problem,'' which involves the difficulty of discriminating the optical transient associated with the gravitational-wave event from other astrophysical transients \citep{MeBe2012,CoBe2015}. In particular, Cowperthwaite and Berger \citep{CoBe2015} recently showed how the existence of pre-existing deep template images in the gravitational-wave sky localization region can greatly improve the detection rate over searches without prior template images. They also showed that kilonovae can be robustly separated from other known and hypothetical types of transients utilizing cuts on color and rise time. 

The remainder of this paper is organized as follows. 
We describe the formalism used in this paper in section \ref{sec:formalism}.  
We discuss the methods used to optimize telescope allocations and demonstrate their application in section \ref{sec:optimization}.  
We conclude with a discussion of topics for further study in section \ref{sec:Conclusion}.

\section{Formalism}
\label{sec:formalism}

\subsection{Definition of variables}

We seek to derive the optimal observing strategy for a single telescope for LIGO/Virgo follow-up observations. 
Telescope time is our limiting resource.
The goal is to develop an observing program that maximizes the probability of finding an associated optical transient that is expected to have some absolute magnitude $M$. 
The apparent magnitude $m$ of the transient is $m = M+\mu+A(\alpha,\delta)$, where
$\mu$ is the distance modulus and 
$A(\alpha,\delta)$ is Galactic extinction, which is the absorption and scattering of electromagnetic radiation by dust and gas between an emitting astronomical object and the observer. 
We denote the ``flux attenuation'' due to extinction by $a(\alpha,\delta)$, which is proportional to $10^{A(\alpha,\delta)}$ and will enter the merit function we will derive below.
In our analysis, we use extinction maps provided by Schlegel et al.\,\citep{ScFi1998}.

We denote the number of detected photons from the object of interest as $N_\textrm{Object} = \phi_\textrm{Object} t$, where $\phi_\textrm{Object}$ is the detectable flux from the object and $t$ is the observation time.
The noise associated with the observation of this object is 
\begin{equation}
\textrm{Noise} = \sqrt{\phi_\textrm{Object} t + n_\textrm{pix} (\phi_S t + \phi_D t + N_R^2)}, 
\end{equation}
where $\phi_\textrm{S}$ is the sky luminosity in the direction of the object within the photometric aperture, $\phi_\textrm{D}$ is the dark current of the detector, $N_\textrm{R}$ is the read-noise of the detector, and $n_\textrm{pix}$ is the number of pixels encompassed in the point spread function.
The signal-to-noise for detection of the transient is 
\begin{equation}
\textrm{SNR} = \frac{N_\textrm{Object}}{\textrm{Noise}} = \frac{\phi_\textrm{Object} t}{\sqrt{\phi_\textrm{Object} t + n_\textrm{pix} (\phi_S t + \phi_D t + N_R^2)}}.
\end{equation}
In what follows, we will ignore observing overheads due to, for example, image readout and telescope slews.
We expect all sources of interest in the advanced detector era to be in the sky-dominated case, except for perhaps a Galactic supernova or other very nearby event, which is a source-dominated case.
In the source-dominated regime, $\SNR \propto \sqrt{\phi_\textrm{Object} t}$, while in the sky-dominated regime, $\SNR \propto \frac{\phi_\textrm{Object}}{\sqrt{\phi_\textrm{Sky}}} \sqrt{t}$.  
For fixed $\phi_\textrm{Object}$ and $\phi_\textrm{Sky}$, the time required to sustain a given SNR scales as $\phi_\textrm{Object}^{-2}$. 
Due to the inverse square law (ignoring additional cosmological dimming for the redshift regime of interest here), $\phi_\textrm{Object}$ scales as $R^{-2}$. 
This means that the time required to sustain a constant SNR for a target absolute magnitude scales as $R^4$. 
Inverting this relation, the distance out to which we can find the desired magnitude scales very slowly with exposure time, as $t^{1/4}$. 

We will assume that there is a given maximum counterpart absolute magnitude $M_\textrm{max}$.
As the initial goal is to detect a transient optical source with some SNR, any exposure time used beyond what is needed to accomplish this is a waste. 
On the other hand, any exposure that does not go deep enough to acheive this is also a failure. 
If we point the followup telescope in some direction and integrate for a time $t_\textrm{field}$, we accumulate electromagnetic counterpart detection probability over all the galaxies in the field of view for which the integration time exceeds $t_\textrm{max}$, i.e. the exposure time needed to detect the brightest plausible counterpart. Each galaxy's counterpart likelihood is presumed to be proportional to its stellar mass or luminosity. The figure of merit for that observation is given by the integrated detection likelihood over all these galaxies. We go deeper in the nearby galaxies, but obviously most of them are out at the edge of the useful detection volume, residing at a distance where we can just barely detect the brightest plausible source. The figure of merit for an observation is then the volume integral of the detection probabilities.
In the following, we will compute the dependence of the time required to achieve a given SNR on a number of quantities.
This time depends on the the distance to the transient and the stellar mass in the direction of the field.

Singer et al. \citep{SiPr2014} and subsequently Berry et al.\,\citep{BeMa2015} explored the directional dependence of the gravitational-wave likelihood $L_\textrm{GW}(\alpha,\delta)$ in great detail. They showed that when both the Hanford and Livingston interferometers are operating, the sky position reconstructions will look like two antipodal islands on opposite sides of the sky, one over North America and one on the opposite side of the Earth. These occur due to degeneracies from the relative positions of the two interferometers. When more detectors are included, in general the sky positions become more tightly constrained.
We now briefly turn our attention to the likely distance posteriors, $L_\textrm{GW}(R)$.
Due to the antenna pattern of gravitational-wave detectors and the unknown inclination angle of the gravitational-wave source, the possible distances for a given gravitational-wave amplitude cover a very broad range. 
Singer et al. \citep{Si2016} show how combining the gravitational-wave distance posteriors with a galaxy catalog leads to significant reduction in the total time required to image a counterpart.

\begin{figure}[t]
 \includegraphics[width=4in]{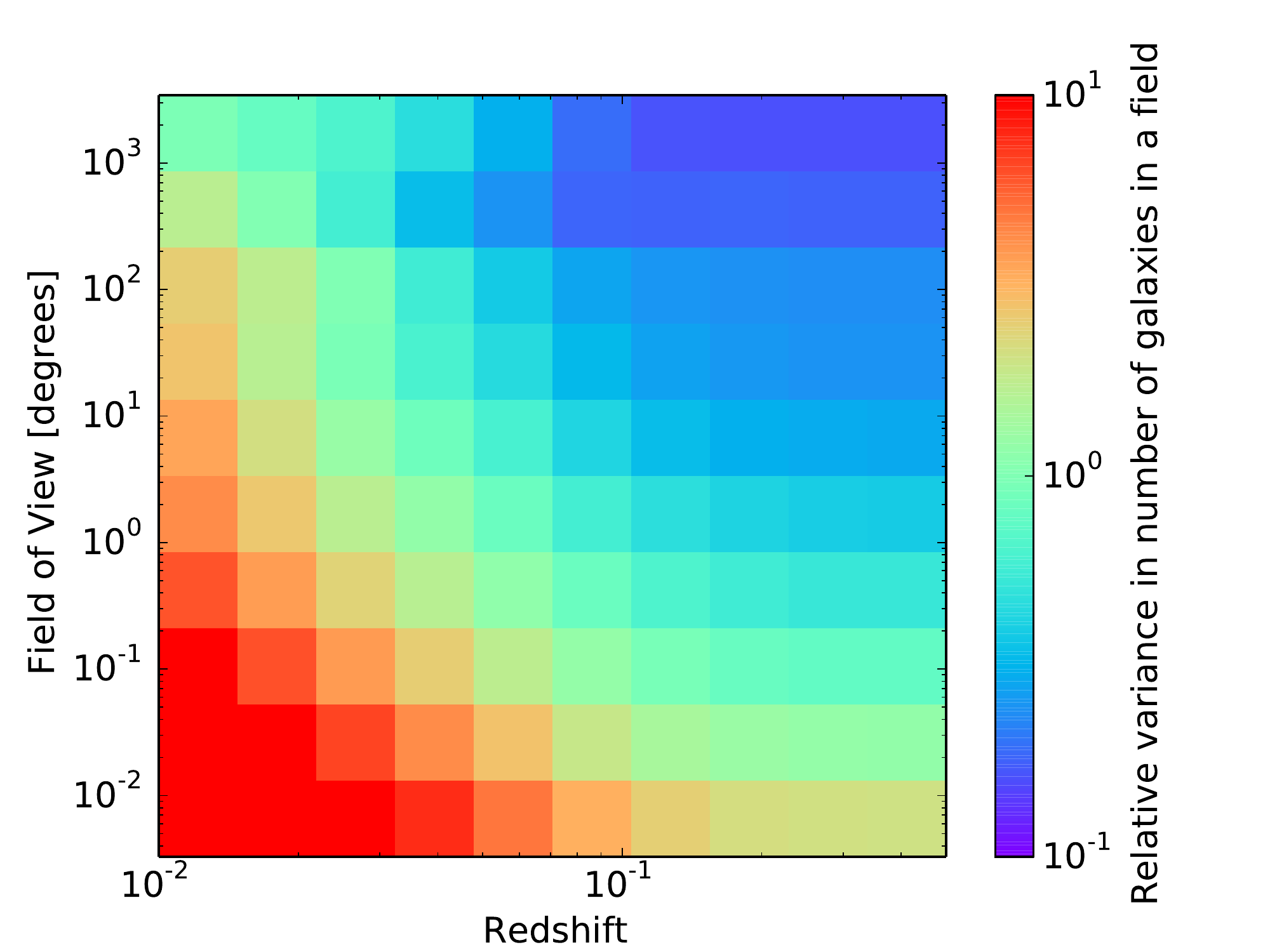} 
 \caption{
   Plot of the relative variance, $\sigma_N / N$, in the number of galaxies as a function of field of view and reach of the gravitational-wave detectors in redshift.
   The results are derived from the 2MASS Photometric Redshift catalog \citep{BiJa2014}. 
   In general, telescopes with a small field of view or when gravitational-wave detectors are sensitive to only nearby transients are likely to have a high variance in the number of galaxies per image.
   On the other hand, telescopes with a large field of view or when the gravitational-wave detectors are very sensitive have little variance from field to field.
 }
 \label{fig:mass}
\end{figure}

We finally explore the assumption about whether a telescope is in the regime where there are many galaxies in a field or few.
This will motivate the examples we use below, where we take one regime where the number of galaxies is significantly different across different fields and the other where it is approximately uniform.
Fig.~\ref{fig:mass} shows the relative variance in the number of galaxies as a function of field of view and reach of the gravitational-wave detectors in redshift.
In general, telescopes with a small field of view or when gravitational-wave detectors are sensitive to only nearby transients are likely to have a high variance in the number of galaxies per image.
On the other hand, telescopes with a large field of view or when the gravitational-wave detectors are very sensitive have little variance from field to field.
It is clear that the regime of interest is significantly dependent on the sensitivity of the gravitational-wave detectors.
As the detectors are commissioned and more detectors enter the network, the sensitivity distance will increase.
Therefore, inclusion of the dependence of the number of galaxies as a function of FOV will be important, and we return to how to incorporate this later.

\subsection{Detection probabilities for fixed telescope time allocations}
\label{sec:probabilities}

We now explain how to derive telescope pointing optimizations given a set of assumptions, which we outline below:
\begin{enumerate}
\item The observations are sky-noise dominated.
\item We know the gravitational-wave likelihood in right ascension ($\alpha$), declination ($\delta$), and distance ($R$). We denote the likelihood as $L_\textrm{GW}(\alpha,\delta,R)$. 
\item We ignore cosmological subtleties like redshift dependence of volume, which is reasonable out to z=0.1. 
\item We assume that the transient arises from an old stellar population (as opposed to young objects) so that the likelihood of a gravitational-wave source in a galaxy is proportional to the galaxy's stellar mass (as opposed to, for example, the star formation rate). 
\item We assume that a particular cadence has been adopted for the fields (i.e. one visit per night, two visits 3 hours apart on the first night, followed by a visit every other night, etc.).
\end{enumerate}
We assume that for each galaxy $i$, there is a probability of detecting a counterpart $p_i$, which can be computed as follows:
\begin{equation}
p_i(t_i)  = \frac{M_i}{M_\textrm{tot}} \frac{L_\textrm{GW}(\alpha_i,\delta_i,R_i)}{L_\textrm{GW tot}} \frac{F_i(t_i)}{a(\alpha_i,\delta_i)}
\label{eq:FOM}
\end{equation}
where $M_i$ is the stellar mass in the galaxy, $a(\alpha_i,\delta_i)$ is the attenuation in the direction, $L_\textrm{GW}(\alpha_i,\delta_i,R_i)$ is the gravitational-wave likelihood in the field, $F_i$ is a weight factor that accounts for assumptions about the luminosity of the counterpart, and $t_i$ is the amount of time allocated to that field. 
For a followup campaign that observes multiple galaxies, the total counterpart detection probability $p_\textrm{tot}$ is the sum over galaxies
\begin{equation}
p_\textrm{tot} = \sum_{i=1}^{N} p_i(t_i).
\end{equation}
If we have multiple fields, the distribution of the total observing time t across field-dependent exposure times $t_i$ is optimal when the partial derivatives of the individual field $p_i$'s with respect to $t_i$ are all equal, such that $\left(\partial p_i(t_i) / \partial t_i\right)=C$. 
This means that when the exposure times are optimized, moving one second of exposure from one field to another has the probability lost in one equal to the probability gained in the other.



\subsection{Electromagnetic Luminosity of gravitational-wave counterparts}

The amount of electromagnetic energy emitted by a coalescence event is not well-known. Not only is the emission mechanism 
poorly understood, but host galaxy extinction will attenuate and redden the light that emerges. 
Cowperthwaite \& Berger \citep{CoBe2015} have summarized both the expected electromagnetic luminoisties and astrophysical contaminants associated with a variety of potential gravitational wave sources. The anticipated peak luminosities for optical counterparts range from $10^{39}$ to $10^{41.5}$ ergs/s, with a considerable range in expected effective temperatures and evolution of the spectral energy distributions across the diversity of astrophysical cataclysm scenarios. If we take a conservative upper bound of $\Phi_\textrm{max}=10^{41}$ ergs/s luminosity as the brightest plausible source, this corresponds to a peak apparent magnitude of $i_\textrm{AB} = 23$ at a distance of 200\,Mpc. These numbers determine $R_0$ and $t_0$, the time needed to detect the brightest plausible transient at a given distance. The LSST exposure time calculator estimates that a single 15 second exposure is sufficient to attain $\textrm{SNR}>5$ at this apparent magnitude, even at 1.5 airmasses in 1'' seeing. So we can adopt $t_0=15\,s$ and nominal $R_0=200$\,Mpc, for an effective telescope diameter equal to LSST, namely 6.5 meters.
Given these uncertainties, in the following, we will explore two regimes of interest: a delta function prior on luminosity and a flat prior on luminosity.

\subsubsection{Delta function prior on luminosity}
We take the object luminosities to be a delta function, $\delta(\phi-\phi_0)$, such that all of the electromagnetic counterpart objects have the same luminosity $\phi_0$.
For a fixed exposure time, there is a threshold distance $R_\textrm{max}$ out to which we can detect the transient of interest. 
$F_i(t_i)$ is a Heaviside step function $\Theta(R_\textrm{max}(t_i) - R_i)$, where any galaxy within $R_\textrm{max}$ is given a weight of 1 and further than $R_\textrm{max}$ is given a weight of 0. 
This implies that
\begin{equation}
p_i = \frac{M_i}{M_\textrm{tot}} \frac{L_\textrm{GW}(\alpha_i,\delta_i,R_i)}{L_\textrm{GW tot}} \frac{\Theta(R_\textrm{max}(t_i) - R_i)}{a(\alpha_i,\delta_i)}
\end{equation}
The time dependence arises through $t_{max,i}$, the time needed to detect the brightest counterpart at a distance $R_i$. For a sky-dominated set of observations this time scales as $R_i^4$.  We now express all detection probabilities relative to 
a fiducial host galaxy at a distance $R_0$ for which it would take a time $t_0$ to achieve a 5$\sigma$ detection of the brightest plausible counterpart. 
We can now make a change of variables, since there is a direct relationship between $R_\textrm{max}$ and t, given by $R_\textrm{max}(t_i) = R_0 (\frac{t_i}{t_0})^{1/4}$.
This means that
\begin{equation}
p_i = \frac{M_i}{M_\textrm{tot}} \frac{L_\textrm{GW}(\alpha_i,\delta_i,R_i)}{L_\textrm{GW tot}} \frac{\Theta(R_0 (\frac{t_i}{t_0})^{1/4} - R_i)}{a(\alpha_i,\delta_i)}
\end{equation}

\subsubsection{Flat prior on luminosity}

In this scenario, we adopt a flat prior for the luminosity (in some electromagnetic detection passband) that emerges from the host galaxy, up to an upper limit, so that $P(\phi) = $C for $\phi < \phi_{max}$:
\begin{equation}
P(\phi>\phi_{0}) = 
\begin{cases}
      1-\frac{\phi_0}{\phi_\textrm{max}}, & \text{if}\ \phi_0 < \phi_\textrm{max} \\
      0, & \text{otherwise}
\end{cases}
\label{eq:phi}
\end{equation}
For the sky-noise-dominated case under consideration here, the 5$\sigma$ point source detectable luminosity (in linear luminosity units rather than magnitudes) 
scales as $1/\sqrt{t}$, so the exposure time needed to reach sources fainter than $\phi_{max}$  is $(\phi_{max}/\phi)^2$ longer than needed to detect
$\phi_\textrm{max}$.
The probability of detecting a source of interest, given the flat prior described above, is then a function of the integration time. 
If we scale all exposure times by the time $t_{max}$ needed to achieve 5$\sigma$ sensitivity to $\phi_{max}$, we can determine the counterpart detection probability as a function of exposure time.
Taken together, this implies that 
\begin{equation}
p_i(t) = L_i M_i (1 - \frac{\phi_0 r^2}{\phi_{max} \sqrt{t}})
\end{equation}
We can draw some initial conclusions at this stage. Obviously, a total integration time that falls short of that needed to detect the brightest possible 
counterpart is not time well spent.  Perhaps most importantly, half the detection probability is for sources brighter than half the maximum. In order to achieve 50\% 
detection probability therefore requires that we integrate for 4 t$_{max}$, {\it i.e.} four times as long as is required to detect the brightest plausible counterpart. Attaining 80\% or 90\% counterpart detection probability, however, requires exposures times of 25 t$_{max}$ and 100 t$_{max}$, respectively.

\section{Optimization}
\label{sec:optimization}

We now explore how to optimize time allocations across potential fields. We first explain how to optimize for the 2 pointings case for both the delta and flat luminosity prior. Thereafter, we generalize to N pointings.

\subsection{2 pointing case}

For intuition purposes, we now explore a situation where we have 2 potential galaxies, with different gravitational-wave likelihoods and masses. In this case, we have 
\begin{equation}
\begin{split}
p_1 = \frac{L_1 M_1 F_1}{a_1} \\ 
p_2 = \frac{L_2 M_2 F_2}{a_2}.
\end{split}
\end{equation}
We now compare the use of the two different luminosity assumptions used in this paper.
In the case of a delta function prior on luminosity,
\begin{equation}
\begin{split}
p_1 = \frac{L_1 M_1}{a_1} \Theta(R_0 (\frac{t_1}{t_0})^{1/4} - R_1) \\ 
p_2 = \frac{L_2 M_2}{a_2} \Theta(R_0 (\frac{t_2}{t_0})^{1/4} - R_2),
\end{split}
\end{equation}
where we constrain the total time allocated to be $t = t_1 + t_2$. 
In this case, one simply allocates time $t$ to the field with the larger $\frac{L M}{R^4}$ until $R = R_0 (\frac{t}{t_0})^{1/4}$, and then switches over to the other.

In the case of a flat luminosity function,
\begin{equation}
\begin{split}
p_1 = L_1 M_1 (1 - \frac{\phi_0 r^2}{\phi_{max} \sqrt{t_1}}) \\ 
p_2 = L_2 M_2 (1 - \frac{\phi_0 r^2}{\phi_{max} \sqrt{t_2}}),
\end{split}
\end{equation}
where we again constrain the total time allocated to be $t = t_1 + t_2$. 
The total probability of detecting a counterpart is simply given by $p_\textrm{tot} = p_1 + p_2$ and $p = L_1 M_1 (1 - \frac{\phi_0 r^2}{\phi_{max} \sqrt{t_1}}) + L_2 M_2 (1 - \frac{\phi_0 r^2}{\phi_{max} \sqrt{t_2}})$. To maximize the probability, we set $\frac{\partial p}{\partial t_1} = 0$, which implies that $\frac{t_1}{t_2} = \left(\frac{M_1 L_1}{M_2 L_2} \right)^{2/3}$.

\subsection{N pointing case}

The extension to an arbitrary number of pointings is straightforward.
In the delta function luminosity case,
\begin{equation}
p_\textrm{tot} = \sum_{i = 1} \frac{M_i}{M_\textrm{tot}} \frac{L_\textrm{GW}(\alpha_i,\delta_i,R_i)}{L_\textrm{GW tot}} \frac{\Theta(R_0 (\frac{t_i}{t_0})^{1/4} - R_i)}{a(\alpha_i,\delta_i)}
\end{equation}
where $t = \sum_i t_i$. 
Similar to the above, the pointings are rank-ordered by $\frac{L M}{R^4}$, the time is allocated on the first field until $R = R_0 (\frac{t}{t_0})^{1/4}$, and then switches over to the next, and so on.

In the case of a flat luminosity function,

  \[
    p_i = \left\{\begin{array}{lr}
        \int_0^\infty L_i \rho \Omega (1 - \frac{\phi_{5,i}^{em}}{\phi_{max}}) r^2 dr  &  (1 - \frac{\phi_{5,i}^{em}}{\phi_{max}}) \geq 0\\
        0 & (1 - \frac{\phi_{5,i}^{em}}{\phi_{max}}) < 0
        \end{array}\right\}
  \]

This means that
\begin{equation}
p_i(t) = \rho \Omega \int_0^\infty \frac{L_\textrm{GW}(\alpha_i,\delta_i)}{a(\alpha_i,\delta_i)} (1 - \frac{\phi_0 r^2}{\phi_{max} \sqrt{t}}) r^2 dr
\end{equation}

We now assume the telescope has integrated long enough to see the brightest source in some distant galaxy.
Due to the finite sensitivity of the gravitational-wave detectors, there is also a $R_\textrm{min}$ and $R_\textrm{max}$ to which they are sensitive. 
We make the further assumption that the distance dependence of the gravitational-wave likelihood is largely independent of position across the field of view.
Due to the antenna factors of the gravitational-wave detectors, this can be a poor assumption and improvements will be explored in the future.
We can make this integral more concrete by putting in explicit limits of integration and realizing that the angular portion factors out and we are left with a purely radial integral,

\begin{equation}
p_i(t) = \Omega \frac{L_\textrm{GW}(\alpha_i,\delta_i)}{a(\alpha_i,\delta_i)} \rho \int_{R_\textrm{min}}^{R_\textrm{max}} (1 - \frac{\phi_0 r^2}{\phi_{max} \sqrt{t}}) r^2 dr
\end{equation}
Solving this integral
\begin{equation}
p_i(t) = \rho \Omega L_\textrm{GW}(\alpha,\delta,R) \left(R_\textrm{max}-R_\textrm{min}-\frac{R_\textrm{max}^5 \alpha }{5 \sqrt{t}}+\frac{R_\textrm{min}^5 \alpha }{5 \sqrt{t}}\right)
\end{equation}
Taking the partial derivative of $p_j$ with respect to t,
\begin{equation}
\frac{d p_j}{d t} = \frac{\rho \Omega L_\textrm{GW}(\alpha,\delta,R) \left(R_\textrm{max}^5-R_\textrm{min}^5\right) \alpha }{10 t^{3/2}}
\end{equation}
This means that $t_i \propto L_i^{2/3}$.

\subsection{Demonstration}
\label{sec:demonstration}

\begin{figure}[t]
 \includegraphics[width=3.5in]{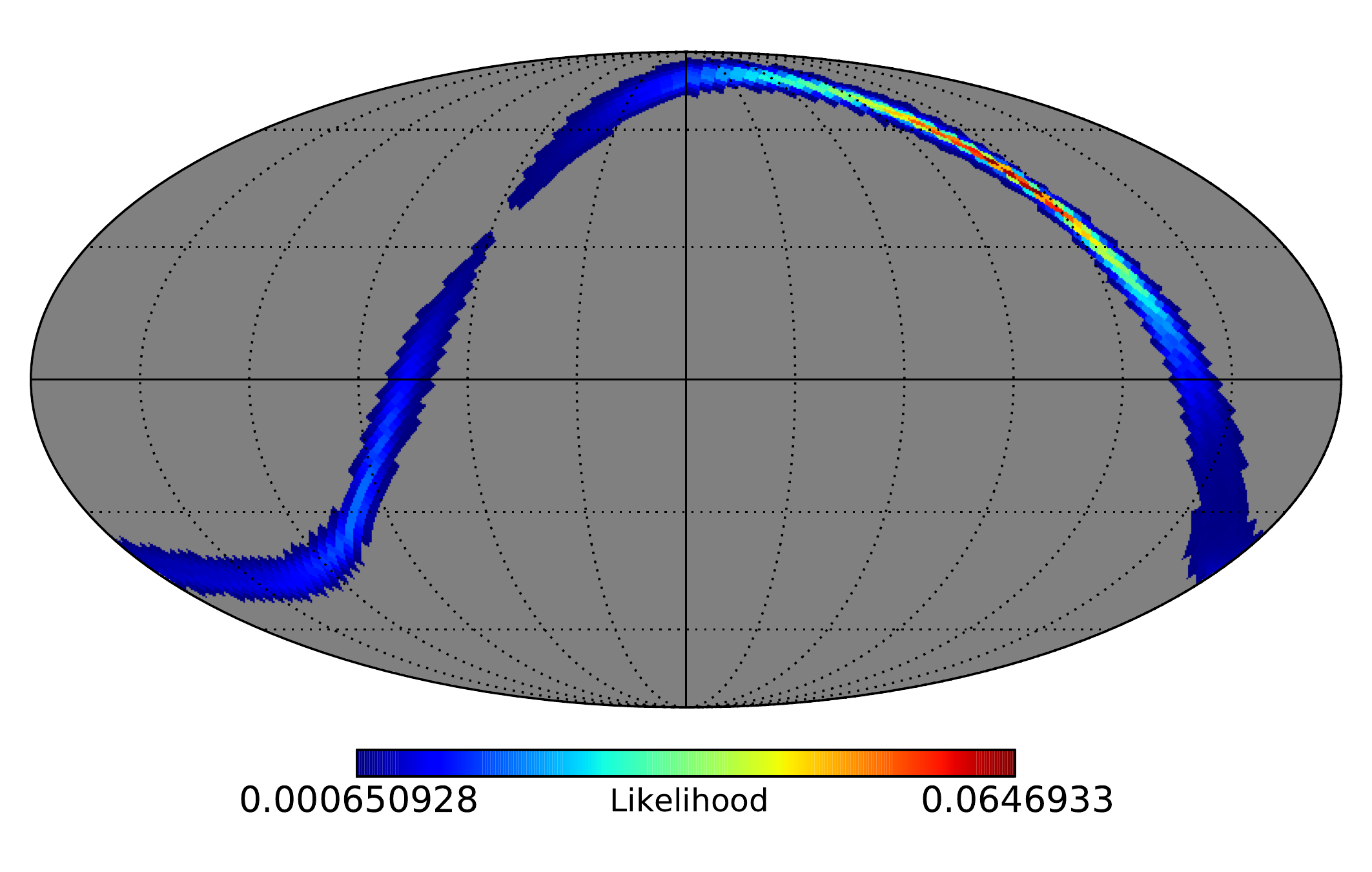} 
 \caption{
   The gravitational-wave likelihood $L_\textrm{GW}(\alpha,\delta,R)$ for the event of interest. This likelihood is to give the scaled optimal observation time allocation for this same event, assuming a continous mass distribution, such that the probability goes as $\left( \frac{L_\textrm{GW}(\alpha,\delta,R)}{a(\alpha,\delta)} \right)^{2/3}$.
 }
 \label{fig:example}
\end{figure}

We now provide a demonstration of the technique from the previous sections.
We begin with the case where we have a single gravitational-wave event with an associated skymap, such as in Fig.~\ref{fig:example}, and desire to know how to pursue optimised follow-up.
We assume that we have been allocated a fixed period of time $t$ on a telescope.

The recipe for construction of the pointing directions and time-allocations are as follows.
Depending on the source model and mass distribution assumptions, the relevant metric from the previous section is computed. 
For example, in the uniform mass density case, the metric is $t_i \propto \left( \frac{L_\textrm{GW}(\alpha,\delta,R)}{a(\alpha,\delta)} \right)^{2/3}$ for the skymap of interest.
The FOVs are rank-ordered by this metric and images taken with time for the allocation appropriate for that field.

\begin{figure*}[t]
 \includegraphics[width=3.5in]{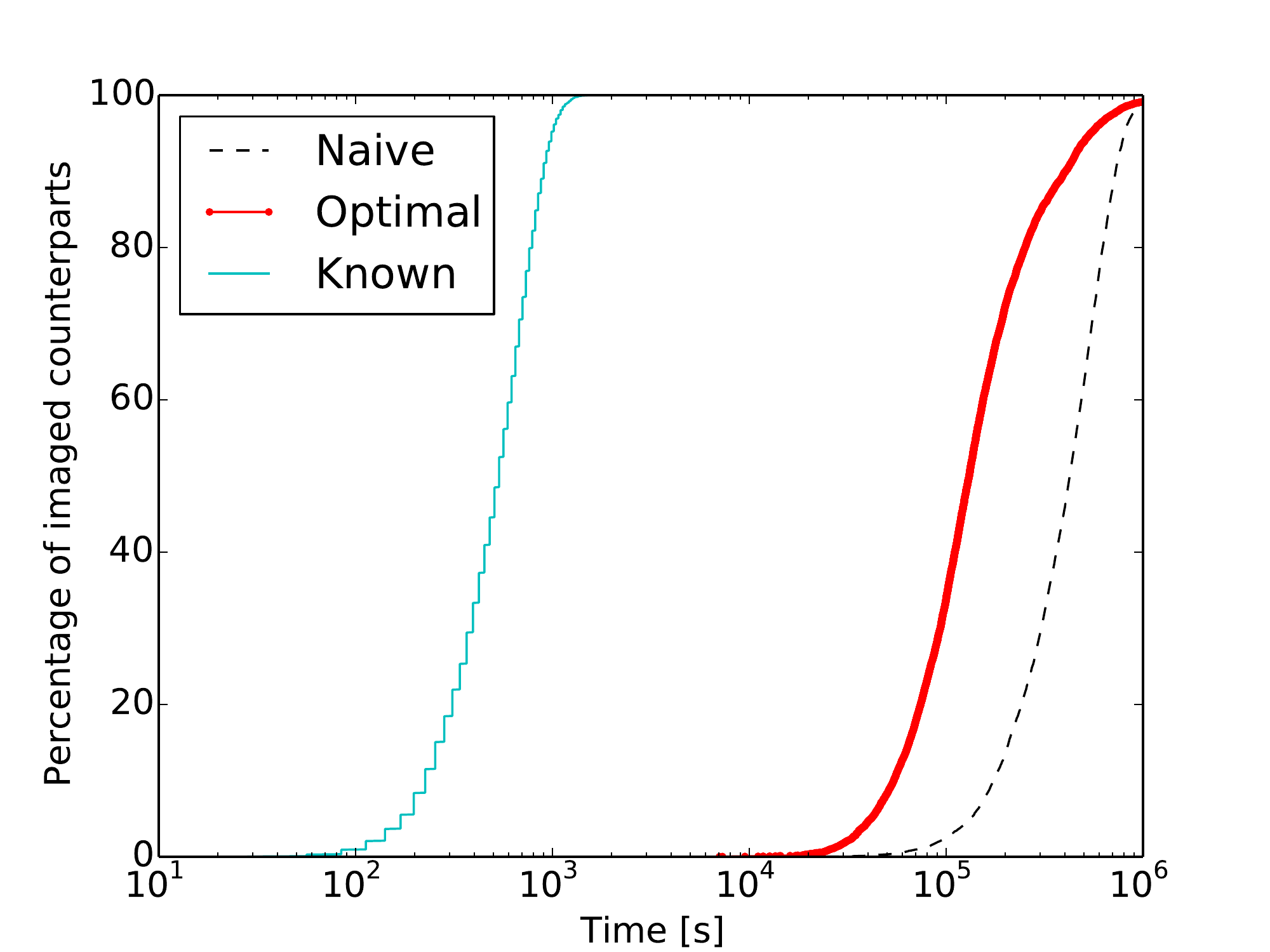} 
 \includegraphics[width=3.5in]{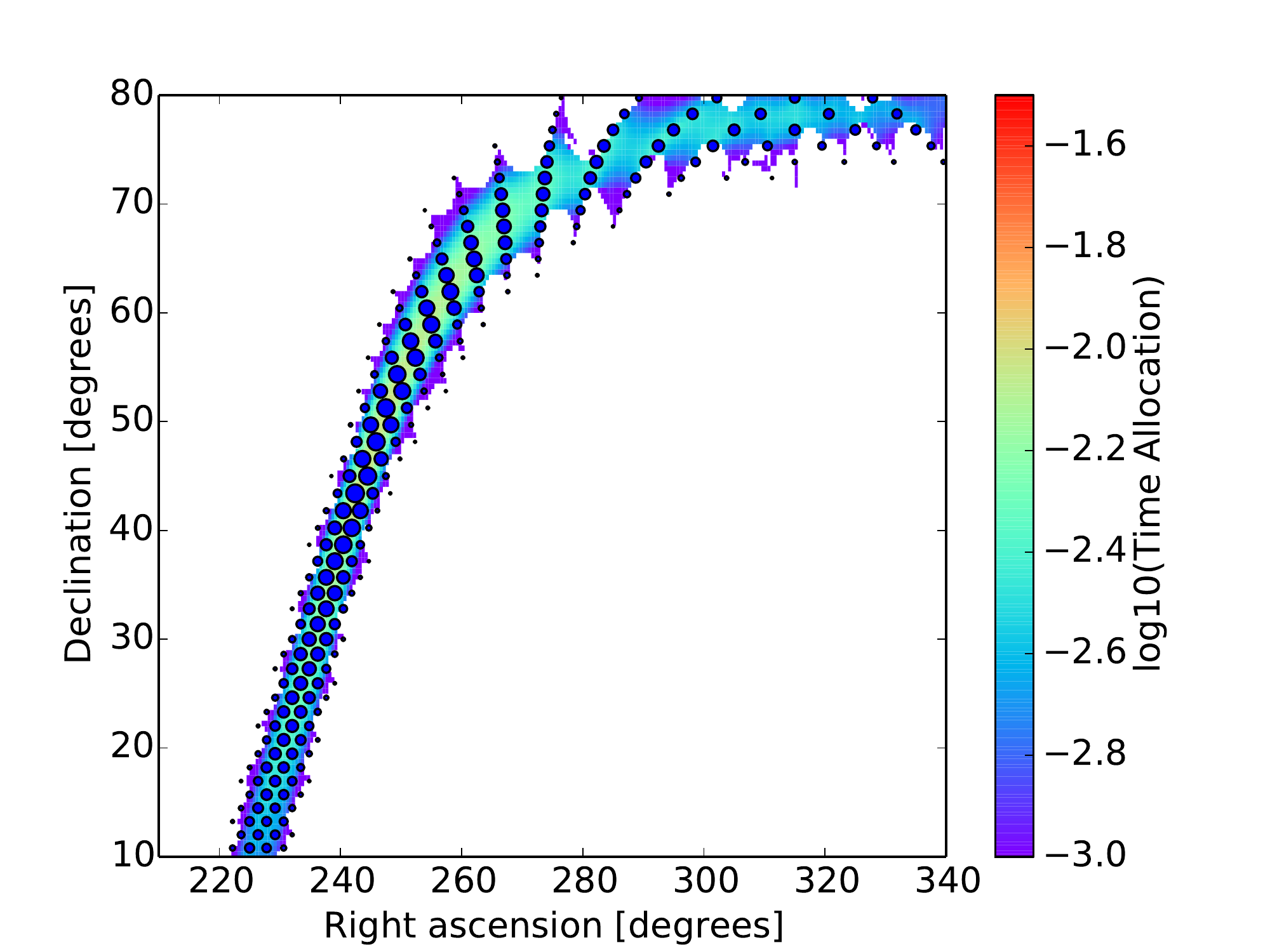} 
  \caption{
   On the left is the proportion of imaged counterparts as a function of time allocated to an event.
   We plot this for three scenarios: where the event location is previously known, where the likelihood is scaled in the optimal way derived in this work, and finally a naive strategy of imaging all fields equally. 
   On the right are the proposed fields for observations by Pan-STARRS using time scalings shown by the size of the dots.
 }
 \label{fig:allocations}
\end{figure*}

We now perform a Monte Carlo simulation where we place sources on the sky consistent with the given skymap and determine the number of images required to successfully recover them.
For concreteness, we adopt the parameters for a source with an absolute magnitude of m=-11.
We adopt as our current telescope the Panoramic Survey Telescope and Rapid Response System (Pan-STARRS) is a telescope designed to discover new Near Earth Objects (NEOs), as well as provide astrometry and photometry of already detected objects.
It has a $3^\circ$ FOV with a limiting magnitude of about 24, taking images of the entire sky about 4 times per month.
We compare three scenarios.
The first is where the event location is previously known. 
In this particular case, we obtain a 50\% detection probability after 540\,s of integration.
In the case where a naive strategy is employed, where all fields are tiled equally, the event can be imaged with 50\% probability in 118\,hrs.
Finally, in the case where the optimal strategy is employed, the event can be imaged in 36\,hrs with 50\% probability.
This corresponds to approximately a factor 3 typical reduction in the amount of time required to image the event.

\section{Conclusion}
\label{sec:Conclusion}

We have described an implementation of an optimization strategy for the detection of gravitational-wave optical counterparts.
We showed how an implementation of this kind can improve searches for these transients.
We find that by making assumptions about the event rate and sky localization abilities of the gravitational-wave detectors, follow-up imaging can be significantly improved.
Therefore, this approach may provide further opportunities for improving electromagnetic follow-ups.

In the future, we can consider the case that we have many events that will be available for follow-up over the science run. 
If all goes well, gravitational-wave detectors will detect an event rate of $N_\textrm{events} = 40$ events per year.
It is possible that there are significantly fewer events than this, if pessimistic astrophysical models prove to be the case.
Luckily, during any given science run, it will quickly become apparant the number of triggers being generated by the gravitational-wave detectors is not as expected. 
Therefore, the number of events assumed can be updated based on the number of events seen in the first few months, for example.

Further, we can explore the coordination of multiple telescopes with different fields of view and limiting magnitudes. 
We expect that a similar formalism can be compiled for this case.
This will be important especially for coordinating, for example, Pan-STARRS and ATLAS. 
Pan-STARRS has a FOV of about 3 square degrees.
Coupled with the reduction in FOV due to the fill factor and hexagonal tiling due to using a circular field, it usually takes multiple exposures.
ATLAS, on the other hand, is 5.4 x 5.4 square degrees.
Due to its larger field of view, no fill factor, and square footprint tiles with very little loss, the difference in the number of images required for an ATLAS and Pan-STARRS field can be about an order of magnitude.
Therefore, an optimal observing strategy must account for this difference.

\section{Acknowledgments}
MC is supported by National Science Foundation Graduate Research Fellowship Program, under NSF grant number DGE 1144152.
CWS is grateful to the DOE Office 
of Science for their support under award DE-SC0007881.
The authors would like to thank Professor Stephen Smartt and Professor John Tonry for comments on the manuscript.

\bibliographystyle{plainnat}
\bibliography{references}

\end{document}